\documentclass[aps, prb, 10pt, twocolumn, superscriptaddress, floatfix]{revtex4-1}

\usepackage{graphicx}
\usepackage{amsmath,amssymb,amsfonts}
\usepackage[usenames, dvipsnames]{color}
\usepackage[breaklinks=false, colorlinks, citecolor=blue, linkcolor=blue, urlcolor=blue]{hyperref}
\usepackage[percent]{overpic}
\usepackage{braket}
\usepackage{bm}
\begin{document}
\title{Quantum versus classical many-body batteries}
\author{Gian Marcello Andolina}
\affiliation{NEST, Scuola Normale Superiore, I-56126 Pisa,~Italy}
\affiliation{Istituto Italiano di Tecnologia, Graphene Labs, Via Morego 30, I-16163 Genova,~Italy}
\author{Maximilian Keck}
\affiliation{NEST,  Scuola Normale Superiore and Istituto Nanoscienze-CNR, I-56126 Pisa,~Italy}
\author{Andrea Mari}
\affiliation{NEST,  Scuola Normale Superiore and Istituto Nanoscienze-CNR, I-56126 Pisa,~Italy}
\author{Vittorio Giovannetti}
\affiliation{NEST,  Scuola Normale Superiore and Istituto Nanoscienze-CNR, I-56126 Pisa,~Italy}
\author{Marco Polini}
\affiliation{Istituto Italiano di Tecnologia, Graphene Labs, Via Morego 30, I-16163 Genova,~Italy}

\newcommand{\red}[1]{\textcolor{red}{#1}}
\newcommand{\blue}[1]{\textcolor{blue}{#1}}

\begin{abstract}
Quantum batteries are quantum mechanical systems with many degrees of freedom which can be used to store energy and that display fast charging. The physics behind fast charging is still unclear. Is this just due to the collective behavior of the underlying interacting many-body system or does it have its roots in the quantum mechanical nature of the system itself? In this work we address these questions by studying three examples of quantum-mechanical many-body batteries with rigorous classical analogs. We find that the answer is model dependent and, even within the same model, depends on the value of the coupling constant that controls the interaction between the charger and the battery itself. 
\end{abstract}

\maketitle

\section{Introduction}

Recently there has been a great deal of interest in quantum batteries (QBs)~\cite{Alicki13, Hovhannisyan13, Binder15, Campaioli17, Ferraro17, Le17,Andolina18,Andolina18b,Campaioli18,Farina18,Julia-Farre18,Zhang18}, i.e.~quantum mechanical systems that are able to store energy. These works have a key common thread in trying to understand whether quantumness yields a temporal speed-up of the charging process. A first, abstract approach~\cite{Binder15,Campaioli17} studied the possibility to charge $N$ systems via unitary operations. The authors introduced a parallel charging scheme, in which each of the subsystems is acted upon independently of the others, and a collective charging scheme, where global unitary operations acting on the full Hilbert space of all subsystems are allowed. In these works it was shown that the charging time scales with $N$, decreasing for increasing $N$. In the collective charging case and for large $N$, the power delivered by a QB is much larger than the one delivered by the parallel scheme. This speed-up was dubbed ``quantum advantage''~\cite{Binder15,Campaioli17,Le17,Ferraro17}. Furthermore, in Ref.~\onlinecite{Campaioli17} it was shown that entanglement is not required to speed-up the evolution of a QB, since states which are confined in the sphere of separable states share an identical speed-up. However, the authors of Ref.~\onlinecite{Campaioli17} pointed out that such highly mixed states host only a vanishing amount of energy, yielding therefore a highly non-optimal result from the point of view of energy storage and delivery. 

\begin{figure}[t]
\centering
\vspace{1.5em}
\begin{overpic}[width=0.9\columnwidth]{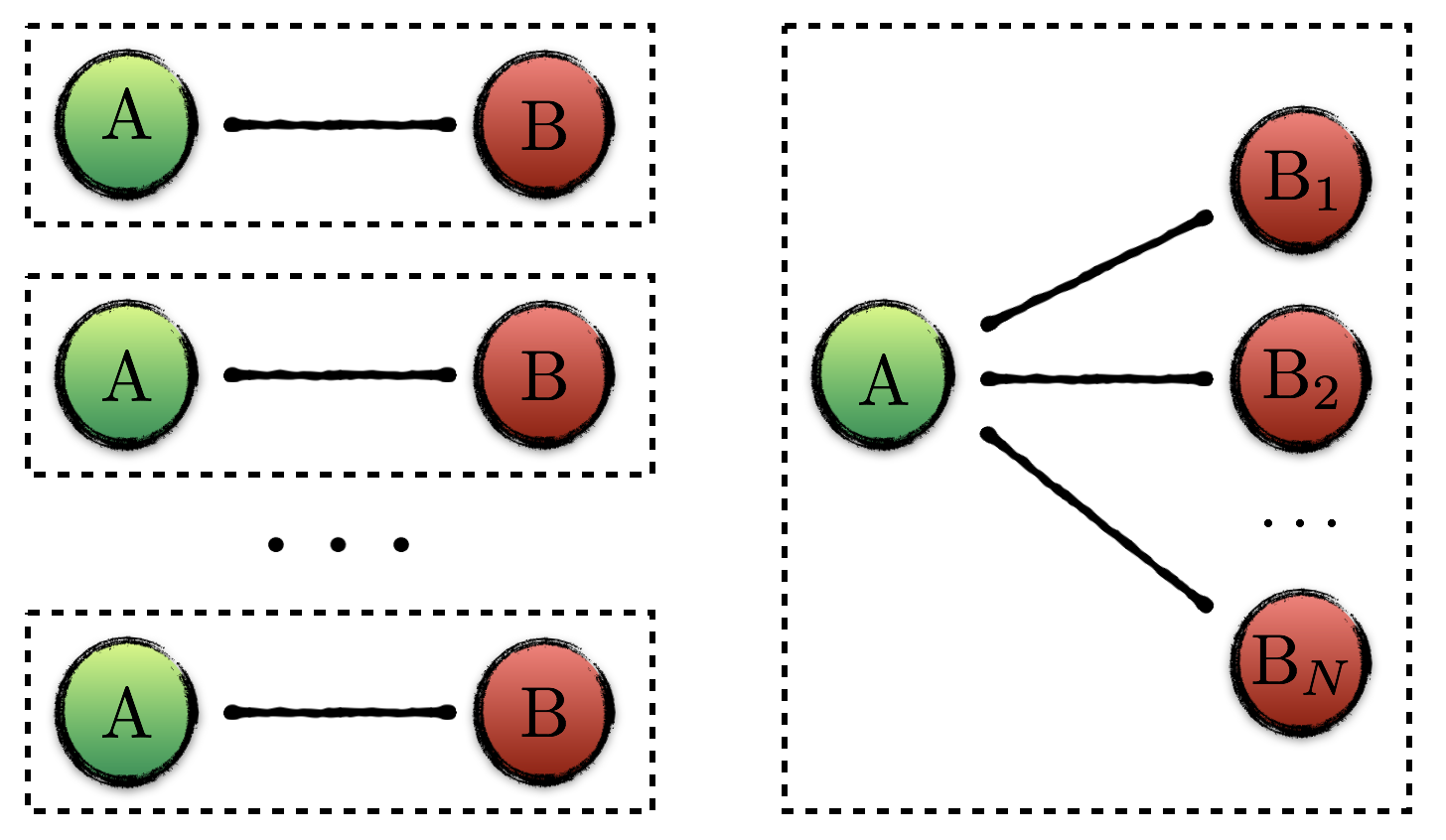}\put(4,31){}\end{overpic}
\caption{(Color online) A sketch of the parallel (left) versus collective (right) charging schemes introduced in the main text.\label{fig:Sketch}}
\end{figure}

In the same spirit, the authors of Refs.~\onlinecite{Ferraro17, Le17,Andolina18,Andolina18b,Farina18} studied similar issues but in realistic setups which can be implemented in a laboratory, such as arrays of qubits in cavities~\cite{Ferraro17,Andolina18,Andolina18b,Farina18} and spin chains in external magnetic fields~\cite{Le17}. In Refs.~\onlinecite{Ferraro17,Andolina18,Andolina18b,Farina18}, the battery units are not charged via abstract unitaries but, rather, by other quantum mechanical systems dubbed ``chargers''. In this framework, the parallel scheme is the one in which each battery is charged by its own charger, independently of the others---see Fig.~\ref{fig:Sketch}. On the contrary, the collective scheme is the one in which all batteries are charged by the very same charger. Also in this context, the collective scheme outperforms the parallel one in terms of speed of the charging process. Finally, the authors of Ref.~\onlinecite{Le17} demonstrated that quantum batteries have the potential for faster charging over their classical counterparts. As they noticed, however, the classical counterparts were assumed to be composed of non-interacting units.

In this Article we compare the performance of QBs with that of their appropriate classical versions. Such comparison is clearly of great interest for foundational reasons but has no implications on the development of scalable solid-state systems where energy transfer processes and their time scales can be studied experimentally. Indeed, any solid-state QB device is going to operate on the basis of electrons, photons, spins, etc, which are inherently described by quantum mechanics. We focus on three models. In the first, a single bosonic mode (the charger) is coupled to $N$ harmonic oscillators (the proper battery composed of $N$ subunits). In the second one, $N$ qubits playing the role of charging units are coupled to another set of $N$ qubits playing the role of the proper battery. Finally, the third one is the Dicke QB introduced in Ref.~\onlinecite{Ferraro17}. In the first case, the performance of  classical and quantum versions of the model is identical. In the second case, the classical version outperforms the quantum one. In the third case, there is a range of values of the charger-matter coupling parameter $g$ for which the quantum (classical) model performs better than the classical (quantum) one. 

Our Article is organized as following. In Sect.~\ref{Comp} we explain how the classical versus quantum comparison is actually carried out in this Article, briefly reviewing the correspondence between quantum commutators and classical Poisson brackets.
In Sect.~\ref{Fom} we recap the charging protocol first introduced in Refs.~\onlinecite{Ferraro17,Andolina18} and introduce the figures of merit needed to evaluate the performance of classical and quantum many-body batteries. In Sect.~\ref{sect:HO} we discuss the first model (single bosonic mode coupled to $N$ harmonic oscillators). We demonstrate analytically that in this case classical and quantum versions of the model display fast charging with the same time scale. In Sect.~\ref{sect:spin} we introduce the second model ($N$ qubits coupled to $N$ qubits) and demonstrate how the classical version of the model outperforms the quantum one. In Sect.~\ref{sect:Dicke} we compare the Dicke QB model introduced in Ref.~\onlinecite{Ferraro17} with the corresponding classical analogue, showing numerically that the relative performance depends on the charger-matter coupling $g$. Finally, in Sect.~\ref{concl} we report a summary of our main findings and our conclusions.

\section{Comparison between quantum and classical mechanics}
\label{Comp}

In quantum mechanics, the evolution of an operator $\hat{O}$ in time $t$ is described by the Heisenberg equation of motion $\hbar~d\hat{O}(t)/dt=i[{\mathcal{H}},\hat{O}(t)]$, where ${\cal H}$ is the Hamiltonian. Moreover, canonically conjugate variables, such as position $\hat{q}_i$ and momentum $\hat{p}_j$, fulfill the commutation relation $[\hat{q}_i,\hat{p}_j]=i\hbar\delta_{i,j}$.
In the case of angular momentum $\hat{\bm J}$, a similar relation holds between different components: $[\hat{J}_i,\hat{J}_j]=\sum_k i\hbar\epsilon_{ijk}\hat{J}_k$, where $\epsilon_{ijk}$ is the Levi-Civita tensor. 

In Hamiltonian mechanics, a classical physical system is uniquely described by a set of canonical coordinates $\boldsymbol{x}^{\rm T}=( {\bm p}, {\bm q})$, where the components $q_i, p_i$ are conjugate variables obeying $\{q_i, p_j\}=\delta_{i,j}$. Here, $\{u,v\}\equiv\sum_i(\partial_{q_i} u~ \partial_{p_i}v- \partial_{p_i}u~ \partial_{q_i}v)$ denotes the Poisson brackets. 

The time evolution of the system is uniquely defined by Hamilton's equations:
    \begin{eqnarray}
  \label{eq:HJ}
\frac{dq_i}{dt}&=&\partial_{p_i}\mathcal{H}^{\rm cl}({\bm x})~, \nonumber \\
\frac{dp_i}{dt}&=&-\partial_{q_i}\mathcal{H}^{\rm cl}({\bm x})~.
\end{eqnarray}

A proper comparison between quantum and classical systems can be made by following the canonical quantization procedure~\cite{DIRAC}.
Once the Hamilton's function $\mathcal{H}^{\rm cl}({\bm x})$ of a classical system is written in terms of conjugate variables with Poisson brackets $\{q_i, p_j\}=\delta_{i,j}$, quantization is carried out by replacing classical coordinates by operators and enforcing canonical commutation relations instead of canonical Poisson brackets.

While finding the classical analog of a quantum system with degrees of freedom that are position and momentum is straightforward and consists in making the replacements $\hat{q}_i\to{q}_i$ and $\hat{p}_j\to{p}_j$, the classical version of quantum mechanical angular momentum is more subtle. It turns out~\cite{BraunBook,Carlos18} that the right choice is to replace the components $\hat{J}_i$ of the angular momentum operator $\hat{\bm J}$, with $\hat{\bm J}^2=\hbar^2J(J+1)$, with the classical canonical coordinates $J_z=J\cos(\theta)$ and $\phi=\arctan(J_y/J_x)$, so that $\{J \cos(\theta), \phi \}=1$, i.e.~$\hat{J}_z \to J\cos(\theta)$, 
$\hat{J}_x \to J\sin(\theta)\cos(\phi)$, and $\hat{J}_y \to J\sin(\theta)\sin(\phi)$. 

In the remainder of this Article we set $\hbar=1$.

\section{Charging protocol and figures of merit} 
\label{Fom}

We start by reviewing a model for the charging process of a QB~\cite{Ferraro17,Andolina18,Andolina18b,Farina18}. As stated above, the classical and quantum cases are both described by an Hamiltonian formalism. We can therefore introduce the charging protocol in terms of a general Hamiltonian, without specifying {\it a priori} whether we treat the classical or quantum case. As such, we will describe the protocol in general, commenting explicitly on the classical and quantum cases only when it is needed.

In our charging protocol~\cite{Ferraro17,Andolina18,Andolina18b,Farina18}, a first system $\rm A$ acts as the energy ``charger''  for a second system $\rm B$, which instead acts as 
the proper battery. They are characterized by local Hamiltonians 
$\mathcal{H}_{\rm A}$ and $\mathcal{H}_{\rm B}$, respectively, which, for the sake of convenience, are both chosen to have zero ground-state energy. We  also assume  ${\rm B}$ to be composed by $N$ non-mutually interacting elements. (Effective interactions between these elements are induced by the charger. In the Dicke QB case, for example, the cavity mode induces effective interactions between all the qubits.) In the quantum case, the system at time $t=0$ is in a pure factorized state  $| \psi\rangle_{\rm A} \otimes |0\rangle_{\rm B}$,  $|0\rangle_{\rm B}$ being the ground state of $\mathcal{H}_{\rm B}$ and $| \psi\rangle_{\rm A}$ having mean local energy $E^{(N)}_{\rm A}(0)\equiv {_{\rm A} \langle} \psi|\mathcal{H}_{\rm A}|\psi\rangle_{\rm A} >0$, where $N$ is the number of elements which compose the battery. Analogously, in the classical case we impose that the system B at time $t=0$ is in the configuration with the lowest energy and we fix the energy in the charger A to be $E^{(N)}_{\rm A}(0)>0$. 

By switching on a coupling Hamiltonian $\mathcal{H}_1$ between A and B, our aim is to provide as much energy as possible to $\rm B$, in some finite amount of time $\tau$, the charging time of the protocol.
For this purpose, we write the global Hamiltonian of the AB system as
\begin{equation}
\label{eq:protocol}
\mathcal{H}(t) \equiv \mathcal{H}_{\rm A}+\mathcal{H}_{\rm B}+\lambda(t)\mathcal{H}_1~,
\end{equation}
where  $\lambda(t)$ is a time-dependent parameter that represents the external
control we exert on the system, and which we assume to be given by
a step function  equal to $1$ for $t\in[0,\tau]$ and zero elsewhere. 
Accordingly, in the quantum case, we denote by $|\psi(t) \rangle_{\rm AB}$ the evolved
state of the AB system at time $t$, its total energy $E(t) \equiv {_{\rm AB}\langle} \psi(t) |\mathcal{H}(t)| 
\psi(t) \rangle_{\rm AB}$ being constant at all times with the exception of the switching points, $t=0$ and $t=\tau$, where some non-zero energy can be transferred to ${\rm AB}$ by the external control. (See Ref.~\onlinecite{Andolina18} for a detailed analysis of the energy cost of modulating the interaction.) 

The same conditions hold in the classical case where we denote by $\boldsymbol{x}^{\rm T}(t)=( \boldsymbol{p}(t), \boldsymbol{q}(t))$ and $E(t)=\mathcal{H}^{\rm cl}\big(\boldsymbol{x}(t)\big)$ the solution of Hamilton's equations of motion and the total energy  at time $t$, respectively. Here, $ \boldsymbol{p}$ and $ \boldsymbol{q}$ are classical conjugate variables. It is also useful to define the vector $\boldsymbol{x}_{\rm B}^{\rm T}(t)=( \boldsymbol{p}_{\rm B}(t), \boldsymbol{q}_{\rm B}(t))$, denoting the position in phase space of B at time $t$.

In the quantum case, we are mainly interested in the mean local energy of the battery at the end of the protocol, i.e.
\begin{equation}\label{stored energy}
E^{(N)}_{\rm B}(\tau)\equiv {\rm tr}[\mathcal{H}_{\rm B} \rho_{\rm B}(\tau)]~,
\end{equation}
$\rho_{\rm B}(\tau)$
being the reduced density matrix of the battery at time $\tau$. It is worth noticing that while $E^{(N)}_{\rm B}(\tau)$ does not
necessarily represent the amount of energy that one can recover from the battery after charging, it has been shown that for 
large enough $N$ this is not a relevant issue~\cite{Andolina18b}.
In the classical case, the corresponding quantity is the energy in B, $E^{(N)}_{\rm B}(\tau)=\mathcal{H}^{\rm cl}_{\rm B}(\boldsymbol{x}_{\rm B}(\tau))$.

The performance of the charger-battery set-up can be studied by analyzing the average storing power 
$P^{(N)}_{\rm B}(\tau)\equiv  E^{(N)}_{\rm B}(\tau)/\tau$. Specifically, we define 
the maximum average power as $\bar P^{(N)}_{\rm B}\equiv \max_\tau [P^{(N)}_{\rm B}(\tau)]$.
Finally, we introduce the optimal charging time $\bar{\tau}$, $\bar P^{(N)}_{\rm B}=P^{(N)}_{\rm B}(\bar{\tau})$, and the energy at the maximum power, $\bar{E}^{(N)}_{\rm B}\equiv{E}^{(N)}_{\rm B}(\bar{\tau})$.

Our aim is to compare the parallel charging scenario against the collective one~\cite{Binder15,Campaioli17,Ferraro17}. As mentioned above, we define as a parallel charging, the protocol in which $N$ batteries are independently charged by $N$ chargers. Each charger has an energy $E_{\rm A}^{(1)}(0)$. Conversely, the  collective charging case is the one in which all $N$ batteries are charged by the same charger. In order to do a clear comparison, in the collective charging case we impose that the charger has total energy equal to the sum of the energies of all the chargers of the parallel charging scheme, i.e.~$E_{\rm A}^{(N)}(0)=NE_{\rm A}^{(1)}(0)$.

Since we are interested in comparing the power of the protocols, we denote by the symbol ${\bar{P}_{\sharp}}$ (${\bar{P}_{\parallel}}$) the maximum power in the collective (parallel) protocol.
Following Ref.~\onlinecite{Campaioli17}, we introduce the so-called collective advantage: 
\begin{equation}\label{Gamma}
\Gamma \equiv \frac{\bar{P}_{\sharp}}{\bar{P}_{\parallel}}~.
\end{equation}
We have $\bar{P}_{\sharp}=\bar{P}^{(N)}_{\rm B}$ and  $\bar{P}_{\parallel}=N\bar{P}^{(1)}_{\rm B}$. The latter property follows from the fact that the power in the parallel charging scheme is trivially extensive.

The figure of merit in Eq.~(\ref{Gamma}) quantifies how convenient is to charge a battery in a collective fashion rather than in a parallel way. While in Refs.~\onlinecite{Campaioli17,Ferraro17} this quantity is named ``quantum advantage'', it is possible to define $\Gamma$ also in the classical case. Since our main goal is to compare quantum and classical batteries, we will denote by $\Gamma_{\rm qu}$ the collective advantage produced by a quantum  Hamiltonian and by $\Gamma_{\rm cl}$ the collective advantage produced by the analog classical Hamiltonian. 
What matters is therefore the ratio
\begin{equation}\label{R}
R\equiv \frac{\Gamma_{\rm qu}}{\Gamma_{\rm cl}}~.
\end{equation}
If $R=1$, the QB and its classical analog share the same collective boost in the charging process. Conversely, having $R>1$ means that there is a genuine quantum advantage. Finally, $R<1$ means that the collective dynamics in the classical model is more beneficial. 

The quantity $R$ will be crucial below in determining if fast charging is due to exquisitely quantum resources or, rather, if it has a collective (i.e.~many-body) origin due to effective interactions between the battery subunits, which is present also in the classical case.

\section{Harmonic oscillator batteries}
\label{sect:HO}

In this Section we study a system composed by $N+1$ harmonic oscillators, one acting as a charger while the remaining $N$ playing the role of the proper battery. This system is described by the following Hamiltonian,
\begin{eqnarray}\label{eq:Hhoho}
\mathcal{H}_{\rm A}&=&\omega_0 a^\dagger a~,\nonumber\\
\mathcal{H}_{\rm B}&=&\omega_0\sum_i b^\dagger_ib_i~,\nonumber\\
\mathcal{H}_{1}&=&g\sum_i \big(a b^\dagger_i+a^\dagger b_i\big)~,
\end{eqnarray}
where $a$ ($b_i$) is the destruction bosonic operator acting on A (on the $i$-th unit of the battery B), and ${\omega}_0$ and $g$ are the characteristic frequency of both systems and the charger-battery coupling parameter, respectively. For simplicity, we choose $E_{\rm A}^{(1)}(0)=\omega_0$.

It is useful to introduce the bright mode~\cite{Ciuti05} $B=\sum_i b_i/\sqrt{N}$,  which is a bosonic mode fulfilling $[B,B^\dagger]=1$. Expressing the Hamiltonian in terms of the bright mode, we obtain:
\begin{eqnarray}
  \label{eq:HhohoB}
\mathcal{H}_{\rm B}&=&\omega_0 B^\dagger B~,\nonumber\\
\mathcal{H}_{1}&=&g_N\left( a B^\dagger+a^\dagger B\right)~,
\end{eqnarray}
where
\begin{equation}\label{eq:g_N}
g_N \equiv \sqrt{N}g~.
\end{equation}
Hence, the AB system is equivalent to two harmonic oscillators with a renormalized coupling $g_N$. It is straightforward to obtain the dynamics of the energy of B, which is independent of the initial state $\ket{\psi}_{\rm A}$  in A. In order to calculate the stored energy~\eqref{stored energy} we find then useful to adopt  the Heisenberg representation, writing $E_{\rm B}(\tau) ={\rm tr}[\rho_{\rm AB}(0)\mathcal{H}_{\rm B}(\tau)]$, where $\rho_{\rm AB}(0)$  is the density matrix of the full system at the initial time, with $\mathcal{H}_{\rm B}(\tau)\equiv e^{i \mathcal{H} \tau} \mathcal{H}_{\rm B}e^{-i \mathcal{H} \tau}$. Expressing $a$ and $b$ as functions of the normal operators $\gamma_{\pm}=(a\pm B)\sqrt{2}$ and using that the latter evolve simply as $\gamma_{\pm}(t) = e^{-i\omega_{\pm}t}\gamma_{\pm}$ with $\omega_{\pm}=\omega_0\pm g_N$, we obtain
 \begin{eqnarray}\label{evolvedH}
\mathcal{H}_{\rm B}(\tau)&=&\frac{\omega_0}{2}\Bigg\{a^\dagger a+B^\dagger B \\&-&\left[\frac{e^{-i2g_{N}\tau}}{2} (a^\dagger a-B^\dagger B+B^\dagger a-a^\dagger B) +{\rm H.c.}\right]    \Bigg\}~, \nonumber
\end{eqnarray} 
and, finally:
\begin{eqnarray}
  \label{eq:Ebho}
E^{(N)}_{\rm B}(\tau)=N\omega_0\sin^2(g\sqrt{N}\tau)~.
\end{eqnarray}
Defining $Y={\max_x}[\sin^2(x)/x]$, the maximum power reads $\bar{P}^{(N)}_{\rm B}=N\sqrt{N}g\omega_0Y$. Accordingly, we have:
\begin{eqnarray}\label{eq:Gammaho}
\Gamma_{\rm qu}=\sqrt{N}~.
\end{eqnarray}
We note that if  $\ket{\psi}_{\rm A}$ is a coherent state, the evolved state $|\psi(t) \rangle_{\rm AB}$ remains factorized at all times~\cite{Andolina18,WallsMilburn2007}. This is an example where the advantage is present, despite the total absence of correlations.

Now we study the classical analog of the quantum model in Eq.~(\ref{eq:Hhoho}), which can be simply obtained by reversing the quantization procedure and substituting quantum commutators with classical Poisson brackets. The corresponding classical Hamiltonian describes a set of coupled springs:
\begin{eqnarray}
  \label{eq:HhohoCl}
  \mathcal{H}^{\rm cl}_{\rm A}&=&\frac{\omega_0}{2}\left(p_a^2+q_a^2\right),\nonumber~\\
\mathcal{H}^{\rm cl}_{\rm B}&=&\frac{\omega_0}{2}\sum_i\left(p_{b_i}^2+q_{b_i}^2\right)\nonumber~,\\
\mathcal{H}^{\rm cl}_{1}&=&g  \sum_i \left( q_a q_{b_i}+p_a p_{b_i}\right)~,
\end{eqnarray}
where $(p_a,q_a)$ are conjugate variables of the charger and $(\boldsymbol{p}_{b_{i}},\boldsymbol{q}_{b_{i}})$ are conjugate variables of the $i$-th battery. As earlier, we choose $E_{\rm A}^{(1)}(0)=\omega_0$. We now introduce $Q_b=\sum_i q_{b_i}/\sqrt{N}$ and $P_b=\sum_i p_{b_i}/\sqrt{N}$. The classical Hamiltonian becomes
\begin{eqnarray}\label{eq:HhohoClR}
\mathcal{H}^{\rm cl}_{\rm B}&=&\frac{\omega_0}{2}\left(P_{b}^2+Q_{b}^2\right)\nonumber~,\\
\mathcal{H}^{\rm cl}_{1}&=&g_N\left( q_a Q_{b}+p_a P_{b}\right)~.
\end{eqnarray}
We conclude that also in the classical case the model maps into that of two coupled oscillators with a renormalized coupling $g_N$. 

Hamilton's equations of motion follow from Eqs.~\eqref{eq:HJ}, \eqref{eq:HhohoCl}, and~\eqref{eq:HhohoClR}:    
\begin{eqnarray}
  \label{eq:hohoHJ}
\frac{d{p}_a}{dt}&=&-\omega_0q_a-g_N Q_b~,\nonumber \\
\frac{d{q}_a}{dt}&=&\omega_0p_a+g_N P_b~,\nonumber \\
\frac{d{P}_b}{dt}&=&-\omega_0Q_b-g_Nq_a~,\nonumber \\
\frac{d{Q}_a}{dt}&=&\omega_0P_b+g_Np_a~.
\end{eqnarray}
Solving these equations we find that, irrespective of the particular initial condition, the stored energy reads $E^{(N)}_{\rm B}(\tau)=N\omega_0\sin^2(g\sqrt{N}\tau)$. This implies
\begin{eqnarray}
  \label{eq:GammahoCl}
\Gamma_{\rm cl}=\sqrt{N}~,
\end{eqnarray}
and $R=1$. This is the main result of this Section. For the case of harmonic oscillator batteries defined in (\ref{eq:Hhoho}), fast charging, i.e.~$\Gamma \propto \sqrt{N}$, is solely due to the collective behavior of the underlying many-particle system, and does not have its roots in the quantumness of its Hamiltonian.

\section{Spin batteries}
\label{sect:spin}

In this Section we study a system composed by $N$ qubits, acting as charger, coupled to another set of $N$ qubits, which play the role of the battery. 
The quantum Hamiltonian is
\begin{eqnarray}\label{eq:HSpins}
\mathcal{H}_{\rm A}&=&\omega_0\left(J^{(a)}_z+\frac{N}{2}\right)~,\nonumber\\
\mathcal{H}_{\rm B}&=&\omega_0\left(J^{(b)}_{z}+\frac{N}{2}\right)~,\nonumber\\
\mathcal{H}_{1}&=&4g \left(J^{(a)}_xJ^{(b)}_x+J^{(a)}_yJ^{(b)}_y\right)~,
\end{eqnarray}
where $J^{(a)}_\alpha$ ($J^{(b)}_\alpha$) with $\alpha=x,y,z$ are the components of a collective spin operator of length $J=N/2$ acting on the Hilbert space of the charger A (battery B), while all the other parameters have the same meaning as in Eq.~\eqref{eq:Hhoho}.

\begin{figure}[t]
  \begin{minipage}[c]{0.5\linewidth}
  \centering
 \begin{overpic}[width=4.45cm]{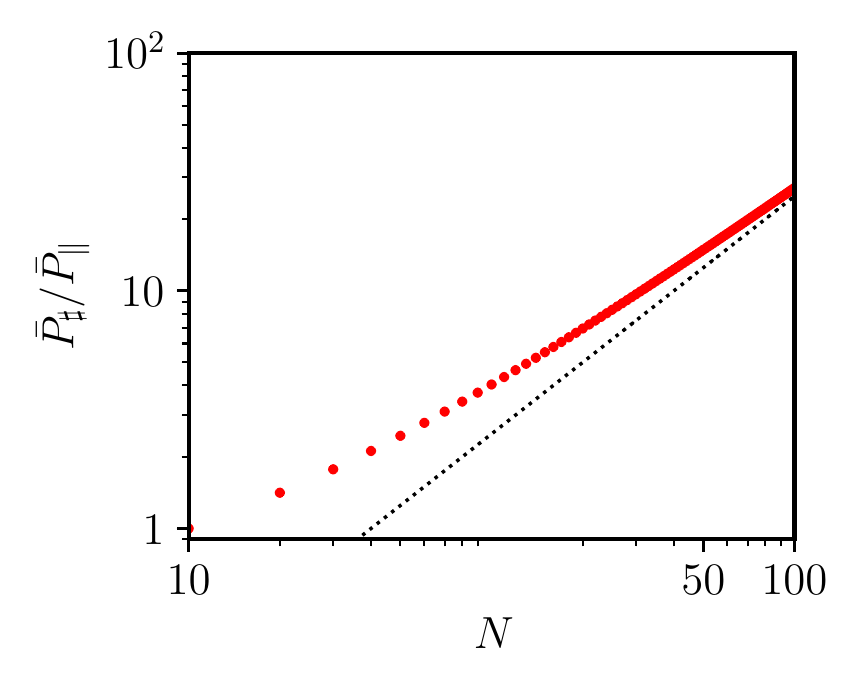}
 \put(1,70){(a)}
 \end{overpic}
  \end{minipage}%%
  \begin{minipage}[c]{0.5\linewidth}
  \centering
 \begin{overpic}[width=4.45cm]{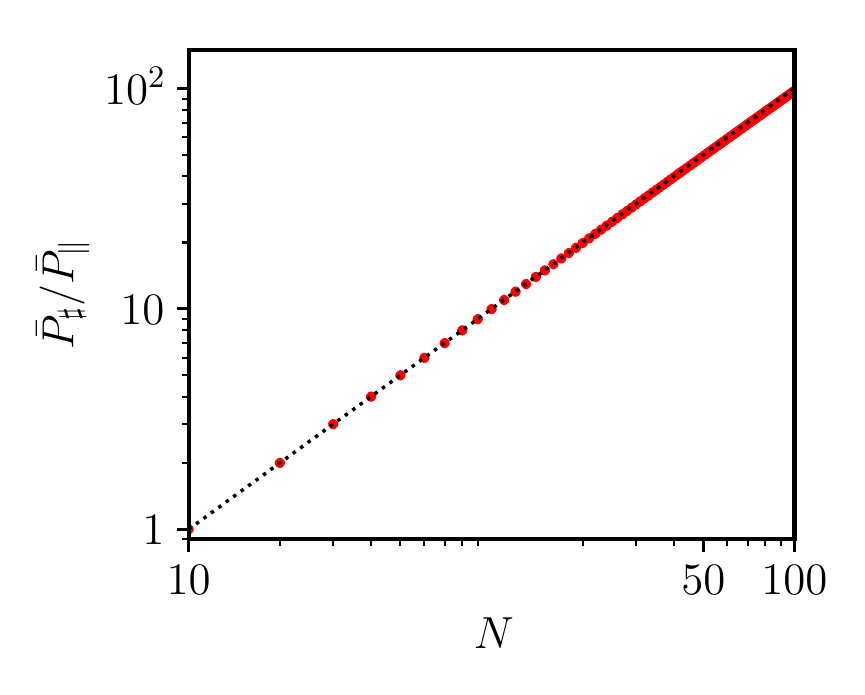}
 \put(1,70){(b)}
 \end{overpic}
  \end{minipage} 
  
     \begin{minipage}[c]{\linewidth}
  \centering
 \begin{overpic}[width=\linewidth]{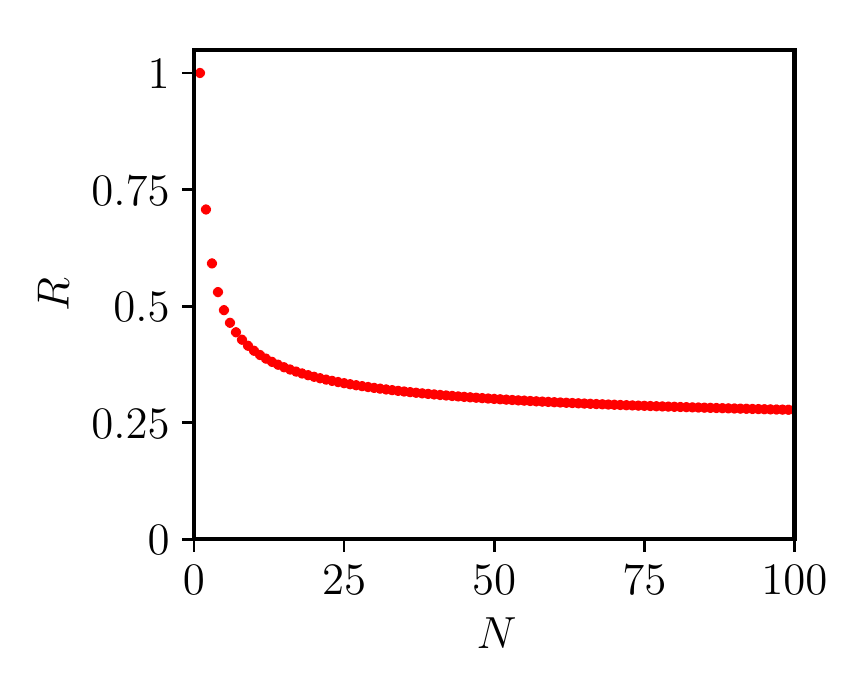}
 \put(1,70){(c)}
 \end{overpic}
  \end{minipage} 
  \caption{(Color online) Performance of quantum and classical spin batteries---see Sect.~\ref{sect:spin}. Panel (a) shows the advantage $\Gamma_{\rm qu}$ in the quantum case, plotted as a function of $N$, in a log-log scale. The black dashed line represents perfectly linear scaling in $N$, i.e.~$\alpha=1$ in Eq.~(\ref{eq:GammaSpins}), with a proportionality constant on the order  of $\sim 0.25$.
  Panel (b) Same as in panel (a), but for the classical case. In this case the scaling is again linear in $N$ with proportionality constant that is, however, equal to $1$. Panel (c) shows the ratio $R = \Gamma_{\rm qu}/\Gamma_{\rm cl}$ as a function of $N$. Notice that, for large enough $N$, $R$ approaches $\sim 0.25$, i.e.~the ratio between the prefactors of the linear scaling with $N$ of the quantum and classical advantages. Results in this figure do not depend on~$g$.\label{fig:3}}
\end{figure}

Defining $\mathcal{H}_{0}=\mathcal{H}_{\rm A}+\mathcal{H}_{\rm B}$, the propagator in the interaction picture simply reads $\tilde{U}_t=e^{i\mathcal{H}_{0}t}e^{-i\mathcal{H}t}=e^{-i\mathcal{H}_1t}$. Hence, in this model there is no dependence of the dynamics on the energy scale $\omega_0$, and $\tilde{U}_t$ depends only on the product $gt$. As in the case of Eq.~(\ref{eq:Gammaho}), this scaling implies that the collective advantage $\Gamma_{\rm qu}$ for this model does not depend on the value of $g$ but only on $N$.
 In Fig.~\ref{fig:3}(a) we report the log-log plot of the collective advantage $\Gamma_{\rm qu}$ as a function of $N$. Fits to the numerical data (not shown) indicate a quasi-linear dependence on $N$ for large $N$ of the form 
\begin{equation}\label{eq:GammaSpins}
\Gamma_{\rm qu}\propto  {N^{\alpha}}~,
\end{equation}
with $\alpha \sim 1$ and a proportionality constant $\sim0.25$.

We now move on to analyze the classical case. Following the discussion of Sect.~\ref{Comp}, we model the 
analog classical Hamiltonian as
\begin{eqnarray}
  \label{eq:HSpinsCL}
    \mathcal{H}^{\rm cl}_{\rm A}&=&N\omega_0\frac{\big[\cos(\theta_a)+1\big]}{2},\nonumber~\\
\mathcal{H}^{\rm cl}_{\rm B}&=&N\omega_0\frac{\big[\cos(\theta_b)+1\big]}{2}\nonumber~,\\
\mathcal{H}^{\rm cl}_{1}&=&g N^2\sin(\theta_a)\sin(\theta_b)\cos(\phi_a-\phi_b)~,
\end{eqnarray}
where  $(N\cos(\theta_a)/2,\phi_a)$ and $(N\cos(\theta_b)/2,\phi_b)$ are conjugate variables~\cite{BraunBook,Carlos18}.

Hamilton's equations of motion follow from Eqs.~\eqref{eq:HJ} and~\eqref{eq:HSpinsCL}. We find
\begin{eqnarray}\label{eq:HSpinsH}
&&\frac{d\cos(\theta_a)}{dt}=2gN \sin(\theta_a)\sin(\theta_b)\sin(\phi_a-\phi_b),\nonumber \\
&&\frac{d\phi_{a}}{dt}=\omega_0-2 g N\cot(\theta_a) \sin(\theta_b)\cos(\phi_a-\phi_b)~.
\end{eqnarray}
Since the Hamiltonian is invariant under the exchange of variables $a\leftrightarrow b$, the equations of motion for $\cos(\theta_{b}) $ and $\phi_{b}$ can be simply obtained by exchanging $a\leftrightarrow b$.

It is now useful to define $\varphi_a=\phi_a+\omega_0 t$ and $\varphi_b=\phi_b+\omega_0 t$, which allow us to write Eq.~(\ref{eq:HSpinsH}) as following:
\begin{eqnarray}\label{eq:HSpinsH2}
&&\frac{d\cos(\theta_a)}{dt} =2gN \sin(\theta_a)\sin(\theta_b)\sin(\varphi_a-\varphi_b),\nonumber \\
&& \frac{d\varphi_{a}}{dt}=-2 g N\cot(\theta_a) \sin(\theta_b)\cos(\varphi_a-\varphi_b)~.
\end{eqnarray}
These equations show that the only energy scale in the problem is $gN$. On the basis of simple dimensional analysis we therefore expect $\bar{\tau}\propto 1/(gN)$. Accordingly, since the energy of the system is extensive, this will yield $\bar{P}_{\parallel} \propto N$ while $\bar{P}_{\sharp}\propto N^2$ leading to 
 $\Gamma_{\rm cl} \propto N$. This argument is not asymptotic, i.e.~does not only apply for $N\gg 1$.
In Fig.~\ref{fig:3}(b) we plot the classical collective advantage obtained by solving numerically Hamilton's equation of motion. Indeed, we clearly see a linear growth in $N$, also for small values of $N$, perfectly consistent with the dimensional argument.

Finally, in Fig.~\ref{fig:3}(c) we show the ratio $R$ defined as in Eq.~(\ref{R}), for the case of our spin batteries. We conclude that, for this model, quantum mechanical dynamics yields a {\it disadvantage} rather than an advantage, as $R<1$ for all $N$. This is the second main result of this Article.

\section{Dicke batteries}
\label{sect:Dicke}

\begin{figure}[t] 
  \begin{minipage}[c]{0.5\linewidth}
  \centering
 \begin{overpic}[width=4.45cm]{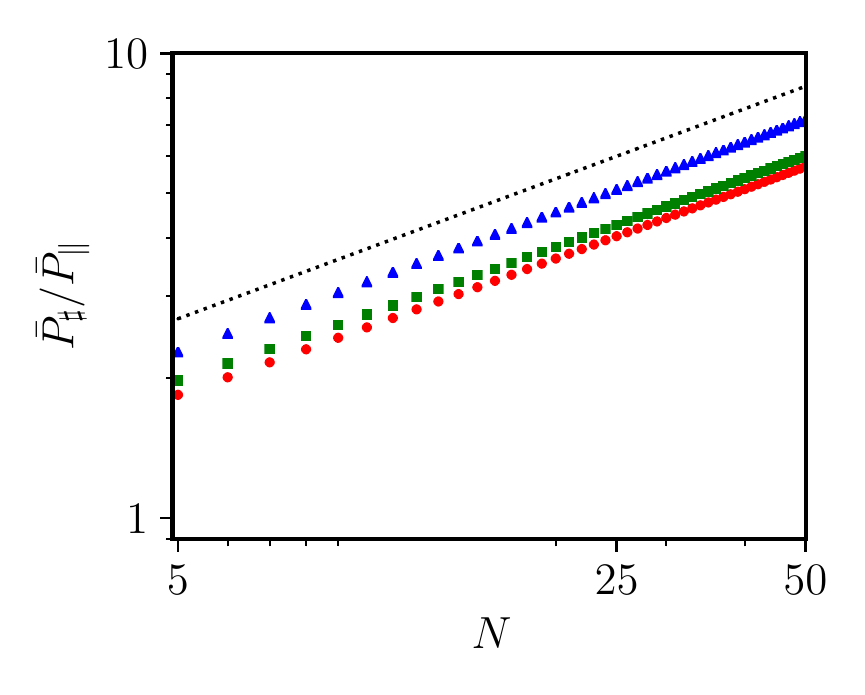}
 \put(1,70){(a)}
 \end{overpic}
  \end{minipage}%%
  \begin{minipage}[c]{0.5\linewidth}
  \centering
 \begin{overpic}[width=4.45cm]{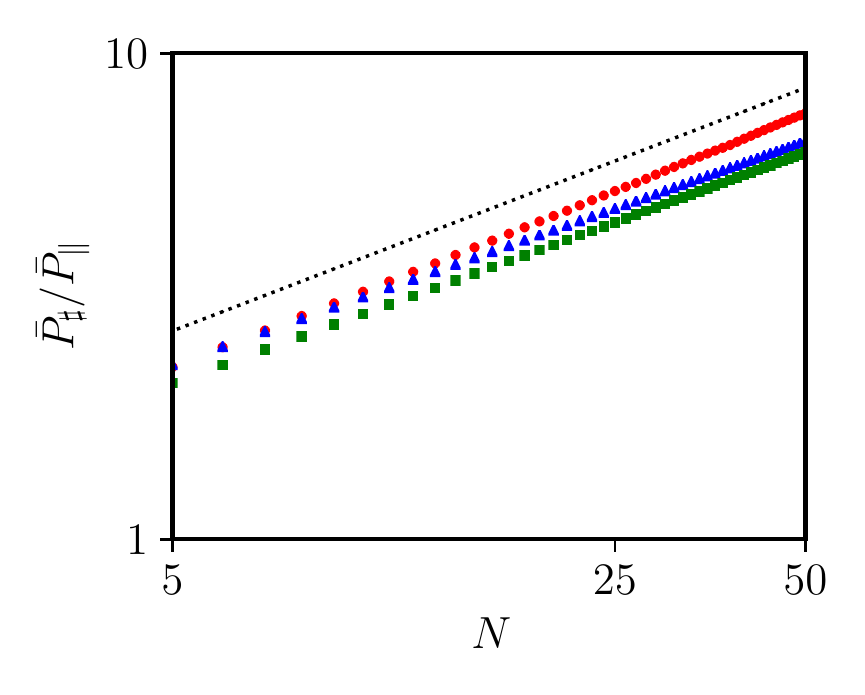}
 \put(1,70){(b)}
 \end{overpic}
  \end{minipage} 
   \begin{minipage}[c]{0.5\linewidth}
  \centering
 \begin{overpic}[width=4.45cm]{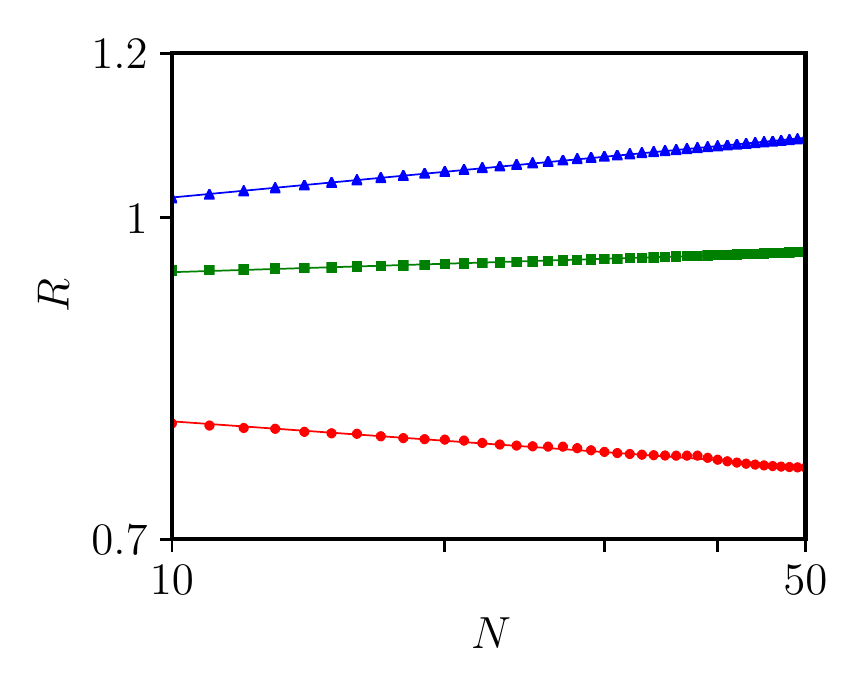}
 \put(1,70){(c)}
 \end{overpic}
  \end{minipage}%%
  \begin{minipage}[c]{0.5\linewidth}
  \centering
 \begin{overpic}[width=4.45cm]{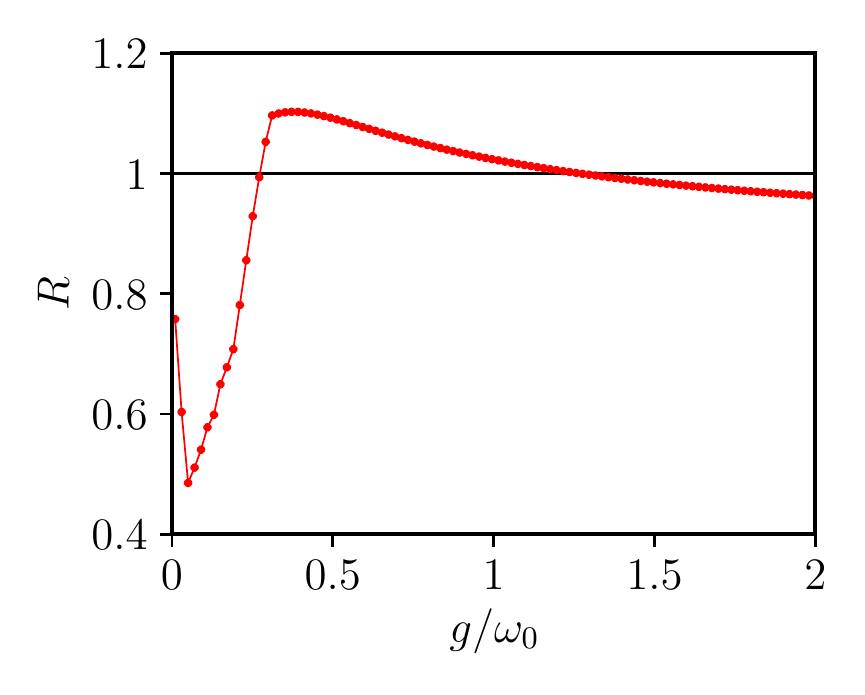}
 \put(1,70){(d)}
 \end{overpic}
  \end{minipage} 
 \caption{(Color online) Performance of quantum and classical Dicke batteries---see Sect.~\ref{sect:Dicke}. Panel (a) shows the advantage $\Gamma$ in the quantum case, plotted as a function of $N$ in a log-log scale. Different symbols refer to different values of the charger-battery coupling parameter $g$: $g=0.01\omega_0$ (red circles), $g=0.5\omega_0$ (blue triangles), and $g=2 \omega_0$ (green squares). The black dashed line represents a scaling of the form $\Gamma_{\rm qu} \propto \sqrt{N}$, i.e.~$\alpha=0$ in Eq.~(\ref{eq:GammaDicke}). Panel (b) Same as in panel (a), but for the classical case. Panel (c) shows the ratio $R$ plotted as a function of $N$, for the same values of $g$ reported in panels (a) and (b). Panel (d) shows the ratio $R$ plotted as a function of $g$, for $N=50$. A quantum advantage on the order of $10\%$  can be observed in a small interval around  $g\simeq 0.5\omega_{0}$.\label{fig:2}}
\end{figure}
In this Section we study the case of Dicke batteries~\cite{Ferraro17,Andolina18b}. In a Dicke QB, one cavity mode, acting as charger, is coupled to $N$ qubits, which play the role of the battery. The quantum Hamiltonian is~\cite{Ferraro17} (see also Refs.~\onlinecite{Dicke54,footnoteDicke})
\begin{eqnarray}
  \label{eq:Dicke}
  \mathcal{H}_{\rm A}&=&\omega_0 ~a^\dagger a~,\nonumber\\
\mathcal{H}_{\rm B}&=&\omega_0\left(J_z+\frac{N}{2}\right)~,\nonumber\\
\mathcal{H}_{1}&=&2g \left(a^\dagger+a\right)J_x~,
\end{eqnarray}
where $J_\alpha$ with $\alpha=x,y,z$ are the components of a collective spin operator of length $J=N/2$, while all the other parameters have the same meaning as in Eq.~\eqref{eq:Hhoho}. As in the other models introduced in previous Sections, we choose $E_{\rm A}^{(1)}(0)=\omega_0$.  Moreover, for the sake of simplicity, we fix $\ket{\psi}_{\rm A}$ to be a Fock state. In Ref.~\onlinecite{Andolina18,Andolina18b} it was shown that the particular choice of the initial state does not change qualitatively the collective advantage. While a detailed analysis of Dicke QBs is reported in Ref.~\onlinecite{Ferraro17}, here we summarize the main findings---Fig.~\ref{fig:2}(a)---and compare them with those obtained for the classical analog of a Dicke QB. 

In Fig.~\ref{fig:2}(a) we plot the collective advantage $\Gamma_{\rm qu}$ of a Dicke QB for different choices of the coupling parameter $g$. In agreement with Ref.~\onlinecite{Ferraro17}, fits to the numerical data (not shown) suggest the following power-law scaling in the limit of large $N$
\begin{eqnarray}\label{eq:GammaDicke}
\Gamma_{\rm qu}\propto \sqrt{N}~.
\end{eqnarray}

We now analyze the classical case. In the literature there is a well-established classical analog of the Dicke model~\cite{deAguiar92,Rodriguez18,Carlos18}, which reads as follow
\begin{eqnarray}\label{eq:DickeCl}
    \mathcal{H}^{\rm cl}_{\rm A}&=&\frac{\omega_0}{2}\left(p_a^2+q_a^2\right),\nonumber~\\
\mathcal{H}^{\rm cl}_{\rm B}&=&N\omega_0\frac{\big[\cos(\theta)+1\big]}{2}\nonumber~,\\
\mathcal{H}^{\rm cl}_{1}&=&g\sqrt{2}Nq_a \sin(\theta)\cos(\phi)~,
\end{eqnarray}
where $(p_a,q_a)$ and $(N\cos(\theta)/2,\phi)$ are classical conjugate variables~\cite{BraunBook,Carlos18}. This Hamiltonian describes a spring coupled to a nonlinear pendulum of length $N$.

We would like to stress that the model defined by Eq.~(\ref{eq:DickeCl}) is not a semi-classical approximation of the quantum Hamiltonian in Eq.~(\ref{eq:Dicke}), but represents instead an intrinsically classical description of a classical spin coupled to a cavity, directly obtainable from classical Hamiltonian mechanics. Our aim is indeed not to approximate the quantum model, but to understand the differences between the quantum and the classical batteries.

As in all previous cases, we choose $E_{\rm A}^{(1)}(0)=\omega_0$. We still have the freedom to choose initial conditions, since the previous condition imposes only the constraint $p_a^2(0)+q_a^2(0)=2N\omega_0$. For the sake of simplicity, we choose $p_a(0)=q_a(0)$. We have checked that other initial conditions do not alter our main conclusions.

From Eqs.~\eqref{eq:HJ} and~\eqref{eq:DickeCl} we find Hamilton's equations of motion for the classical Dicke battery:
\begin{eqnarray}\label{eq:DickeHJ}
&&\frac{d{p}_a}{dt}=-\omega_0q_a-\sqrt{2}Ngq_a\sin(\theta)\cos(\phi)~,\nonumber \\
&&\frac{d{q}_a}{dt}=\omega_0p_a~,\nonumber \\
&&\frac{d\cos(\theta)}{dt}=2\sqrt{2}gq_a\sin(\theta)\sin(\phi),\nonumber \\
&&\frac{d\phi}{dt}=\omega_0-2\sqrt{2}gq_a\cos(\phi)\cot(\theta)~.
\end{eqnarray}
We can rescale these equations in such a way to have $P^{2}_{a}(0)+Q^{2}_a(0)=2$, i.e.~$P_a=\sqrt{N}p_a$ and $Q_a=\sqrt{N}q_a$. We obtain:
\begin{eqnarray}\label{eq:DickeHJT}
&&\frac{d{P}_a}{dt}=-\omega_0Q_a-\sqrt{2}g_NQ_a\sin(\theta)\cos(\phi)~,\nonumber \\
&&\frac{d{Q}_a}{dt}=\omega_0P_a~,\nonumber \\
&&\frac{d\cos(\theta)}{dt}=2\sqrt{2}g_NQ_a\sin(\theta)\sin(\phi),\nonumber \\
&&\frac{d\phi}{dt}=\omega_0-2\sqrt{2}g_NQ_a\cos(\phi)\cot(\theta)~,
\end{eqnarray}
where $g_{N}$ has been defined in Eq.~(\ref{eq:g_N}).
We note that, in these equations, the only parameters with physical dimensions (of energy) are $\omega_0$ and $g_N$. Since $\bar{\tau}$ has physical dimensions of  inverse energy (in our units), the optimal charging time must have the following form:
\begin{equation} \label{DEFEFFE}
\bar{\tau}=\frac{1}{g_{N}}F(\omega_0/g_N)~,
\end{equation}
where $F(x)$ is an unknown dimensionless function. From this expression  we can conclude that, as long as  $F(x)$ does not reach  zero for $x=0$, also in the classical scenario the collective advantange parameter will exhibit 
a $\sqrt{N}$ scaling similar to the one in Eq.~(\ref{eq:GammaDicke}) observed for the quantum counterpart, i.e. $\Gamma_{\rm cl} \propto \sqrt{N}$. 
Indeed, assuming $F(0)\neq 0$, from (\ref{DEFEFFE}) it follows that for large enough $N$ the charging time can be approximated as $\bar{\tau}\simeq F(0)/g_N$ with a $1/\sqrt{N}$ scaling. Accordingly, since  the energy is an extensive quantity, we will have, asymptotically, $\bar{P}^{(N)}_{\rm B}\propto N\sqrt{N}$, which implies $\Gamma_{\rm cl} \propto \sqrt{N}$
as anticipated.
To put this observation on a firmer ground, we resort to numerical integration of Eqs.~\eqref{eq:DickeHJ}. In  Fig.~\ref{fig:2}(b) we plot the collective advantage $\Gamma_{\rm cl}$ as a function of $N$, for different values of $g$. A comparison with the expected $\sqrt{N}$ scaling of $\Gamma_{\rm cl}$ in the large-$N$ limit is also shown. (The expected saturation to the $\sqrt{N}$ scaling law requires $g_{N}/\omega_{0}\gg1$ and is therefore difficult to reach numerically for small values of $g/\omega_{0}$.)

We now proceed with a more quantitive comparison between $\Gamma_{\rm qu}$ and $\Gamma_{\rm cl}$. In  Fig.~\ref{fig:2}(c) we report the plot of 
the quantity $R$ of Eq.~(\ref{R}) as a function of $N$, for different values of $g$. We clearly see that the ratio $R$ can be smaller or larger than unity depending on the value of $g$. This is emphasized in Fig.~\ref{fig:2}(d), where we show $R$ as a function of $g$ for $N=50$. This is the third main result of this Article. The quantum advantage shown by a Dicke QB in a window of values of $g$ is  on the order of $10\%$ and therefore not spectacular but clearly indicates the possibility to engineer more complex quantum Hamiltonians to achieve much better quantum performances. These will be the subject of future work. 

\section{Summary and conclusions}
\label{concl}

In this Article we have compared three quantum battery models against their rigorous classical versions in order to better understand the origin of the fast charging phenomenon discussed in previous literature.

In particular, we have defined a genuine {\it quantum advantage} (i.e.~$R>1$) via the ratio $R$ in Eq.~(\ref{R}) between the collective advantages in the quantum and classical cases, $\Gamma_{\rm cl}$ and $\Gamma_{\rm qu}$, respectively.

In the case of harmonic oscillator batteries---see Sect.~\ref{sect:HO}---$R=1$ for all values of $N$ and $g$. Quantum harmonic oscillator batteries defined as in Eq.~(\ref{eq:Hhoho}) do not therefore display any quantum advantage. The case of spin batteries, discussed in Sect.~\ref{sect:spin}, is even worse. In this model, indeed, $R<1$ for all values of $N$ and $g$. 

We can safely conclude that, in these two cases, fast charging in the quantum case (i.e.~the fact that $\Gamma_{\rm qu}$ increases for increasing $N$) is solely due to the collective behavior of the many-body systems described by the quantum Hamiltonians in Eqs.~(\ref{eq:Hhoho}) and~(\ref{eq:HSpins}), which is also present in the corresponding classical Hamiltonians.

The case of Dicke batteries, discussed in Sect.~\ref{sect:Dicke}, is far more richer. In this case, the ratio $R$ depends on the charger-battery coupling parameter $g$ and, for each fixed $N$, can be larger than unity in a range of values of $g$. As evident from Figs.~\ref{fig:2}(c) and~(d), the quantum advantage displayed by a Dicke quantum battery at optimal coupling is on the order of $10\%$. More work is needed to discover quantum models of batteries with larger values of $R$.

For the sake of completeness, we note that the authors of Ref.~\onlinecite{Julia-Farre18} have very recently proposed to study the evolution of the battery state in the energy eigenspace of the battery Hamiltonian. Combining this geometric approach with bounds on the power, they are able to distinguish whether the quantum advantage in a charging process stems either from the speed of evolution or the non-local character of the battery state.

\section{Acknowledgments}
Part of the numerical work has been performed by using the Python toolbox
QuTiP2~\cite{QuTip}. We wish to thank D. Farina, D. Ferraro, P. Erdman, M. Esposito, and, especially, M. Bera, V. Cavina, S. Juli\'a-Farr\`e, and M. Lewenstein for many useful discussions. 


\begin{thebibliography}{77}
\bibitem{footnoteDicke}
The Hamiltonian in Eq.~\eqref{eq:Dicke} has been widely used in the literature~\cite{Carlos18,Julia-Farre18} with a different normalization of the coupling constant, namely with the replacement $g\to g/\sqrt{N}$. This different choice guarantees well-defined results~\cite{Julia-Farre18} if one works in the thermodynamic limit defined by $N\to \infty$, $L\to \infty$ with $n \equiv N/L = {\rm const}$, where $L$ is the length of the cavity. In this limit, the length of the cavity scales with the number $N$ of qubits in order to keep the density $n$ of qubits constant. Whether one uses Eq.~\eqref{eq:Dicke} or Eq.~\eqref{eq:Dicke} with the replacement $g\to g/\sqrt{N}$ ultimately depends on the experimental setup. For example, in circuit-QED setups like the one realized in Ref.~\onlinecite{Fink09}, the length of the photonic cavity (i.e.~the length of the transmission line resonator) does not scale with the number of qubits (i.e.~the number of transmons). Indeed, in Ref.~\onlinecite{Fink09} the resonator is $\sim 20~{\rm mm}$ long, while a transmon has a linear size which is on the order of $300~{\rm \mu m}$. This implies that the resonator used in the setup of Ref.~\onlinecite{Fink09} can host something like $N = 40$-$50$ qubits, without any need to scale its length with $N$ for their accommodation. The authors of Ref.~\onlinecite{Fink09} used the same Hamiltonian as in Eq.~\eqref{eq:Dicke} and Ref.~\onlinecite{Ferraro17} to explain their data.
\end{thebibliography}
\end{document}